# Elliptical Galaxies in Cluster Cores: Luminosity Functions and Selection Procedures

Emilio Molinari

*Osservatorio Astronomico di Brera, via E. Bianchi 46, I-22055 Merate (LC) - Italy*

**Abstract.**
We present here the results about the determination of the Luminosity Function (LF) for elliptical galaxies in the core of fairly rich clusters as a function of redshift and try to compare with the known LF of close, better studied clusters of galaxies. We then outline the first use of a new method of selecting member elliptical galaxies using an unsupervised Artificial Neural Network (ANN) working on a purely photometric set of data.

## 1. Introduction

We present here a preliminary study of the LF of the galaxies in clusters which is a powerful tool for the determination of the past dynamical history of the cluster phenomena (Binggeli et al., 1988, Biviano et al., 1994, Lopez in this conference). Our data span a range in redshift up to $z \sim 0.5$ and the photometric catalog is well suited for the selection and measurements of the elliptical galaxies properties, as it will be shown at the end with a test experiment using a fully unsupervised ANN.

## 2. Observations and Photometry

The photometric catalogs are based on a series of observing runs carried out at the ESO Chilean site of La Silla, using the 3.6 m diameter telescope equipped with the EFOSC camera and a CCD detector which ensured a field of view of $3 \times 5$ arcmin, with a sub-seeing spatial resolution. We adopted the $g$, $r$, $i$ Gunn photometric system and obtained three-color magnitudes for 11 clusters, 7 of which have measured redshift, ranging from 0.15 to 0.50. Some analysis of the data and the catalogs are found in Molinari et al., (1990, 1994) who quote as typical error for the magnitude estimate less than 0.1 mag at $r = 22$.

We used the Inventory package (West and Kruszewski, 1981) for automatic recovering of the position of the objects over a fixed threshold, chosen to be 1% of the sky background. Isophotal magnitudes and radii, fixed aperture colors are then computed. The crowded fields of cluster cores limit the efficiency of the searching algorithm. A test performed introducing $\sim 3000$ model stars yelded $\sim 100\%$, $50\%$ and $\sim 0\%$ efficiencies at $r = 22, 24, 26$ respectively. A second major problem encountered using Inventory was the presence of very extended objects (D and cD galaxies) which leads to a unreliable measurement for the close-by fainter galaxies. We devised a mathematical model which, on the basis of a few term Fourier series, allow to subtract the light contribution of such large objects. The model allows for the presence of any ellipticity and even small decentering of the inner isophotes.

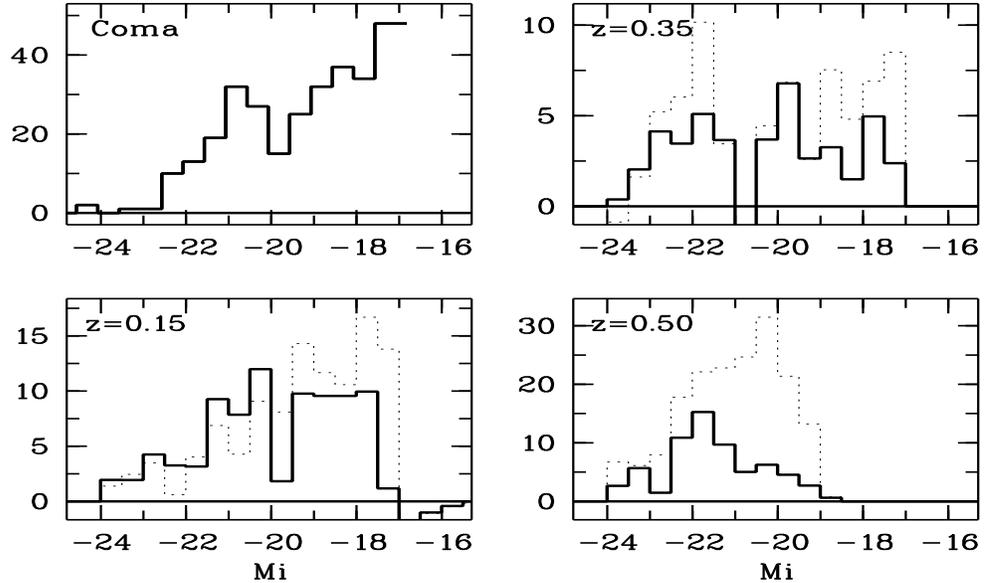

Figure 1. E galaxy LF as a function of redshift. Top left: the Coma cluster LF by Biviano et al. (1994); lower left: the composite LF of 3 clusters at $z \sim 0.15$; top right: one cluster at $z = 0.35$; lower right: the composite LF of two clusters at $z \sim 0.5$.

## 3. Galaxy selection and cluster membership

The color-magnitude (C-M) planes are first used in order to select the elliptical population of the clusters. The well known C-M relation for the elliptical galaxies is well clear in all our plots and a strip 0.5 mag wide was centered on the visible ridge, thus defining as E galaxies those falling within the boundaries. As discussed and analysed by Buzzoni et al. (1993), a color based selection of ellipticals can be reliable and used for testing cosmological changes. Statistical correction for the fore- and background objects was computed dividing each field of view in an *inner* and *outer* part, and estimating the cluster core magnitude distribution as the difference of those in the two regions, properly scaled for the different sampled areas.

## 4. The E-type Luminosity Function

The LF of the selected sample after correction for the non-member contamination shows the presence of gaps in the distribution which do not allow any simple fit with a Schechter or a gaussian function. The colors of such gaps correspond very well with those expected for elliptical galaxies as seen at their relevant redshift, after allowing for k and evolutionary corrections. Furthermore, when comparing with the Virgo cluster LF by Binggeli et al. (1988), the position of the main gap closely fit the magnitude ($M_i = -20$) where the ellipticals change regimes and the dwarf galaxies begins to be the majority of the early types population.

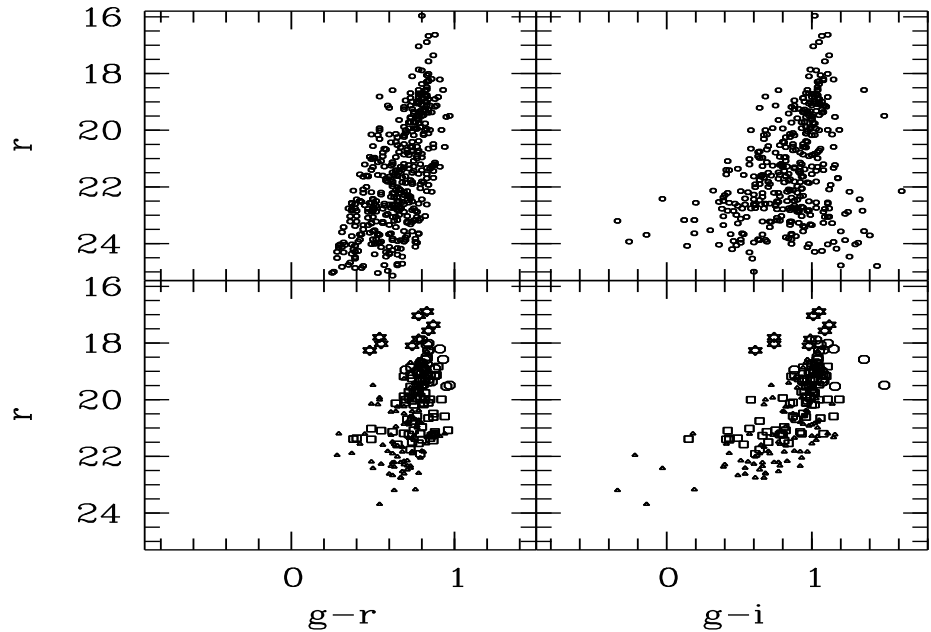

Figure 2. Performance of ANN selection of E galaxies: in the upper two panels the usual stripe selection is shown in the two C-M planes; in the lower panels the E galaxies as identified by ANN are shown in the same C-M planes.

Fig. 1 shows the LFs for all our clusters, grouped in redshift ranges around 0.15, 0.35 and 0.50. The magnitude is the absolute $M_i$, corrected for distance, k and evolutionary correction appropriate for early type galaxies with a single initial burst of star formation (Buzzoni 1989). The comparison with recent estimate of the virialized Coma cluster LF by Biviano et al. (1994) shows an amazing coincidence about the position of the absolute magnitude bin with a low number of galaxies. Biviano et al. interpret the absence of galaxies as the giant ellipticals being less scattered in magnitude and brighter than the corresponding population in a less virialized cluster such as the case of the Virgo cluster. Ongoing merging and/or cannibalism in our cluster is assured by the presence of very extended objects and in one case of a central cD galaxy.

The behaviour at higher redshift looks analogue, with statistically significant (S/N $\geq 2\sigma$) gaps. The brighter magnitudes of these gaps makes intriguing the understanding of a mechanism which involves different sizes of objects along the cosmic time sequence.

## 5. Unsupervised ANN classifies galaxies

One major point in the above discussion is the reliability of the selection of early type galaxies being cluster members on the basis of purely photometric data. We have up to now used the C-M plane with fair results, but this involved only two out of the four measurements we have for our objects. The approach to multi-dimensional clustering of data lead us to explore the possibility of using an ANN, working on a pilot cluster of ~450 objects with complete photometry in the three colors. We chose (after some encouraging results by Smareglia et al., 1994) our ANN to have the topology of a Self Organizing Map (SOM) which achieves the double effect of a natural and automatic clustering of the input 4-dimensional vectors (built with our four photometric properties) and a spatial ordering of a 2-dimensional map, which is an easy tool for the separation of the classes of objects. The SOM, trained with a random subset of 50 objects extracted from the catalogue, was thus able to arrange all the objects in an 160×96 neuron matrix. A natural partition of the plane, cut in two by one of its major diagonals, divided the catalogue in two distinct sets, *on the basis of all the four parameters g, g-r, g-i* and the isophotal radius $R_r$. The comparison of the old method, which rigidly limits the $g$-$r$ vs $r$ plane while allowing a large scatter in the $g$-$i$ vs $r$ plot, with the selection operated by ANN is clearly shown in Fig. 2, where the lower panels present a well defined C-M relation for the early type galaxies in *both* the color planes.

**Acknowledgments.** The setting up and running of the ANN has been carried out in collaboration with R. Smareglia (Osservatorio Astronomico di Trieste).

## Discussion

*M. D'Onofrio*: Just a comment. I am happy of your result on the LF of elliptical galaxies in cluster cores, because we found exactly the same dichotomy between giants and ordinary ellipticals in a volume limited sample of galaxies in Virgo.